\documentclass[aps,prapplied,amsmath,amssymb,twocolumn,superscriptaddress]{revtex4-2}

\usepackage[percent]{overpic}
\usepackage{graphicx}
\usepackage{adjustbox}
\usepackage{fixltx2e}
\usepackage{makecell}
\usepackage{hyperref}
\usepackage{cleveref}
\usepackage{subcaption}
\graphicspath{ {./pictures/} }

\begin{document}


\title{Surface Optimization of Superconducting Aluminum Resonators for Robust Quantum Device Fabrication}


\author{S. J. K. Lang}
\email{simon.lang@emft.fraunhofer.de}
\affiliation{Fraunhofer Institut für Elektronische Mikrosysteme und Festkörpertechnologien EMFT, Munich, Germany}
\author{I. Eisele}
\affiliation{Fraunhofer Institut für Elektronische Mikrosysteme und Festkörpertechnologien EMFT, Munich, Germany}
\affiliation{Center Integrated Sensor Systems (SENS), Universität der Bundeswehr München, Munich, Germany}
\author{A. Maiwald}
\affiliation{Fraunhofer Institut für Elektronische Mikrosysteme und Festkörpertechnologien EMFT, Munich, Germany}
\author{E. Music}
\affiliation{Fraunhofer Institut für Elektronische Mikrosysteme und Festkörpertechnologien EMFT, Munich, Germany}
\author{L. Schwarzenbach}
\affiliation{Fraunhofer Institut für Elektronische Mikrosysteme und Festkörpertechnologien EMFT, Munich, Germany}
\author{C. Moran Guizan}
\affiliation{Fraunhofer Institut für Elektronische Mikrosysteme und Festkörpertechnologien EMFT, Munich, Germany}
\author{J. Weber}
\affiliation{Fraunhofer Institut für Elektronische Mikrosysteme und Festkörpertechnologien EMFT, Munich, Germany}
\author{D. Zahn}
\affiliation{Fraunhofer Institut für Elektronische Mikrosysteme und Festkörpertechnologien EMFT, Munich, Germany}
\author{T. Mayer}
\affiliation{Fraunhofer Institut für Elektronische Mikrosysteme und Festkörpertechnologien EMFT, Munich, Germany}
\author{R. N. Pereira}
\affiliation{Fraunhofer Institut für Elektronische Mikrosysteme und Festkörpertechnologien EMFT, Munich, Germany}
\author{C. Kutter}
\affiliation{Fraunhofer Institut für Elektronische Mikrosysteme und Festkörpertechnologien EMFT, Munich, Germany}
\affiliation{Center Integrated Sensor Systems (SENS), Universität der Bundeswehr München, Munich, Germany}

\date{\today}
\begin{abstract}
Aluminum (Al) remains the central material for superconducting qubits, and considerable effort has been devoted to optimizing its deposition and patterning for quantum devices. However, post-processing strategies focused on oxide removal of niobium (Nb) and tantalum (Ta) -based resonators using buffered oxide etch (BOE), which can not be used for Al. This challenge becomes particularly relevant for industry-scale fabrication with multi-chip bonding, where delays between sample preparation and cooldown require surface treatments that preserve low dielectric loss during extended exposure to ambient conditions.
In this work, we investigate surface modification approaches for Al resonators subjected to a 24-hour delay prior to cryogenic measurement. Passivation using self-limiting oxygen and fluorine chemistries was evaluated utilizing different plasma processes. Remote oxygen plasma treatment reduced dielectric losses, in contrast to direct oxygen plasma. A fluorine-based plasma process was developed that passivated the Al surface for subsequent BOE treatment. However, the fluorine content in the surface resulted in higher loss, identifying fluorine as an unsuitable passivation material for Al resonators. Above all, selective oxide removal using HF (hydrogen fluoride) vapor and phosphoric acid yielded median dielectric losses as low as $\Tilde{\delta}_\mathrm{LP} = 5.7 \times 10^{-7}$ ($Q_\mathrm{LP} \approx 1.7\,\mathrm{M}$) with $\Tilde{\delta}_\mathrm{TLS} = 3.6 \times 10^{-7}$ ($Q_\mathrm{TLS} \approx 2.8\,\mathrm{M}$) in the single photon regime.
Selective oxide removal provides a promising pathway for robust Al-based qubit fabrication, as it preserves low dielectric losses for a 24-hour delay before cooldown.
\end{abstract}
\maketitle

\section{Introduction}
After decades of research, aluminum (Al) remains a widely used material for superconducting transmons. While niobium (Nb) and tantalum (Ta) start to replace Al in the fabrication of resonators due to their favorable properties~\cite{tuokkola_methods_2025,bland_2d_2025}, Al continues to be indispensable for Josephson junction electrodes, making it the most critical component of superconducting qubits. Substantial efforts have therefore been devoted to optimize Al deposition and patterning for superconducting quantum circuits \cite{biznarova_mitigation_2024,fritz_optimization_2019}. Post-processing strategies for Nb and Ta resonators using buffered oxide etch (BOE)\cite{verjauw_investigation_2021, altoe_localization_2022, zheng_nitrogen_2022} typically aim to improve resonator quality with surface cleaning and oxide removal \cite{bafia_oxygen_2024,marcaud_low-loss_2025, crowley_disentangling_2023}. However, those techniques can not easily be applied to Al metallization. Systematic investigations of post-treatments for Al resonators \cite{mahuli_improving_2025, zikiy_high-q_2023, chang_eliminating_2025} have proposed different strategies to address existing limitations, which are particularly relevant because dielectric losses associated with native and process-induced oxides on aluminum or silicon can limit coherence in superconducting circuits \cite{gao_experimental_2008,muller_towards_2019}.

Industrial-scale device fabrication introduces additional constraints. Multi-chip packaging, scheduling, and cleanroom throughput often prevent the immediate transfer of resist stripped samples into a cryostat. Consequently, surface treatments must remain effective even when a cooldown is delayed by several hours. Therefore, the present work focuses on surface treatments applicable to Al resonators with a controlled delay of approximately 24\,h before cooldown, reflecting conditions encountered in scalable fabrication environments.

The main challenge addressed in this study is modifying the Al surface to reduce dielectric losses without damaging the Si surface in the resonator trench. Oxygen and fluorine were selected as promising passivation species. Oxygen-based treatments are compatible with the naturally forming Al$_2$O$_3$ layer, and resist ashing is known to reduce dielectric losses by removing polymer residues \cite{quintana_characterization_2014}. Low-energy plasma oxidation enables controlled growth and saturation of the surface oxide \cite{lang_aluminum_2023} without introducing ion-induced damage or surface roughness that could degrade resonator performance.

Fluorine-based passivation was also explored, motivated by the higher bond energy of Al--F compared to Al--O ($\mathrm{G}_{\mathrm{AlF}} = 675$\,kJ/mol vs.\ $\mathrm{G}_{\mathrm{AlO}} = 502$\,kJ/mol \cite{rumble_crc_2022}), which suggests protection against re-oxidation \cite{kim_atomic_2022,lee_chemical_1990,chen_cf4_2023}. The protection of the Al surface was demonstrated by the repellency of BOE, which normally attacks Al. Prior to fluorination, native oxides were removed to prepare a pristine Al surface for the fluorine plasma treatment. To date, there are no studies showing the influence of fluorine passivation on dielectric losses of superconducting Al resonators.

Finally, selective chemical post-etching was employed to remove SiO$_2$ and Al$_2$O$_3$ independently. HF (hydrogen fluoride) vapor etching selectively removed SiO$_2$ without affecting Al$_2$O$_3$, whereas phosphoric acid exhibited the opposite selectivity. Since oxides are known to host two-level systems (TLS)~\cite{martinis_decoherence_2005, muller_towards_2019, gao_experimental_2008}, removing them is expected to reduce dielectric losses.

Our study demonstrated that a well designed surface preparation with plasma and etch processes enhance the performance of Al-based superconducting resonators for quantum applications.

\section{Analytical Methods}
\label{sec:methodic}
X-ray photoelectron spectroscopy (XPS) measurements quantify elemental composition of surface layers. With in-situ Ar ion milling to remove the upper layers, the technique also enables depth profiling and helps to eliminate contamination such as carbon or native oxides. In this study, we used XPS to analyze the Al, F, and O composition of CF$_4$ plasma-treated Al chip surfaces. All samples were etched for 1\,min with in-situ Ar ion milling prior to the XPS measurement to remove environmental carbon contamination. The Al 2p spectrum was analyzed with respect to the different bonds Al--O (74.5\,eV), Al--O--F (75.8\,eV) and Al--F (76.5\,eV) \cite{Fan_improvements_2007, Oh_improving_2025, Ramos_cleaning_2007}. By fitting these bond peaks to the Al 2p core profile, the individual peak areas can be extracted and define the bonds concentration similar to \cite{An_Impact_2025, chen_cf4_2023}. The bond concentration is calculated relative to the total Al 2p profile area. Additionally, atomic force microscopy (AFM) was used to measure the root-mean-square roughness (R$_\mathrm{RMS}$) of the open silicon surface. Identical areas of 5\,$\mu m^2$ were used for proper comparison. 

\begin{figure}[htbp]
\centerline{\includegraphics[width=0.99\linewidth]{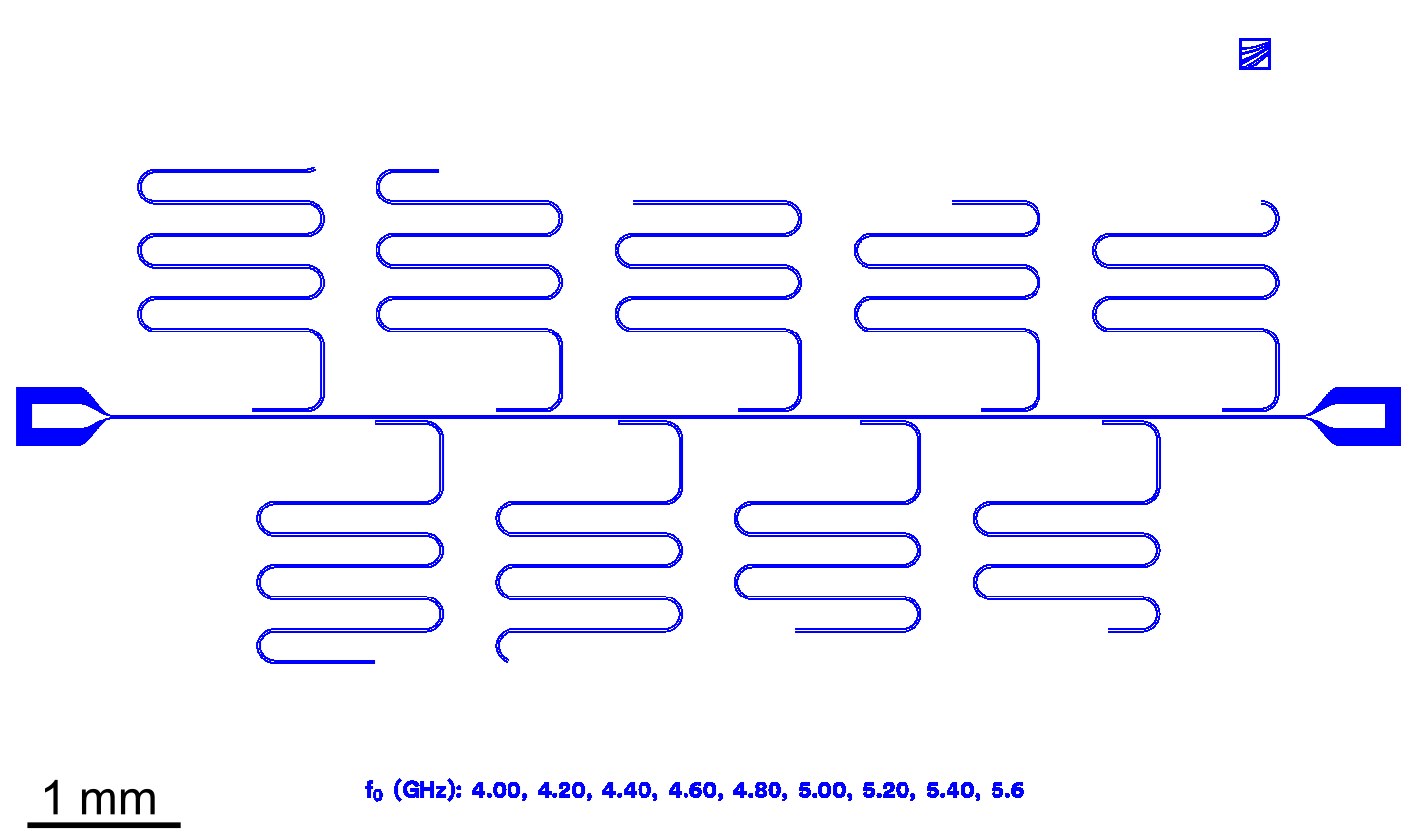}}
\caption{Design of the resonator chips, containing nine CPW meanders with frequencies between 4\,GHz and 5.6\,GHz.}
\label{fig:design_resonator}
\end{figure}

Cryogenic benchmarking of the post-treatments was performed at temperatures below 25\,mK in a Bluefors dilution refrigerator and a cryostat from Qinu. The chip design (see Fig.~\ref{fig:design_resonator}) consists of nine coplanar waveguides (CPW) resonators (gap: 6\,µm, width: 10\,µm) with evenly spaced frequencies between 4\,GHz and 5.6\,GHz, all capacitively coupled to a common feedline. The ground plane surrounding the resonators is patterned with vortex-pinning holes \cite{mcrae_materials_2020}. The cross-section of the resonators as well as the participation-ratio simulation for our geometry is shown in \cite{lang_optimizing_2025}.

\begin{figure}[htbp]
\centerline{\includegraphics[width=0.95\linewidth]{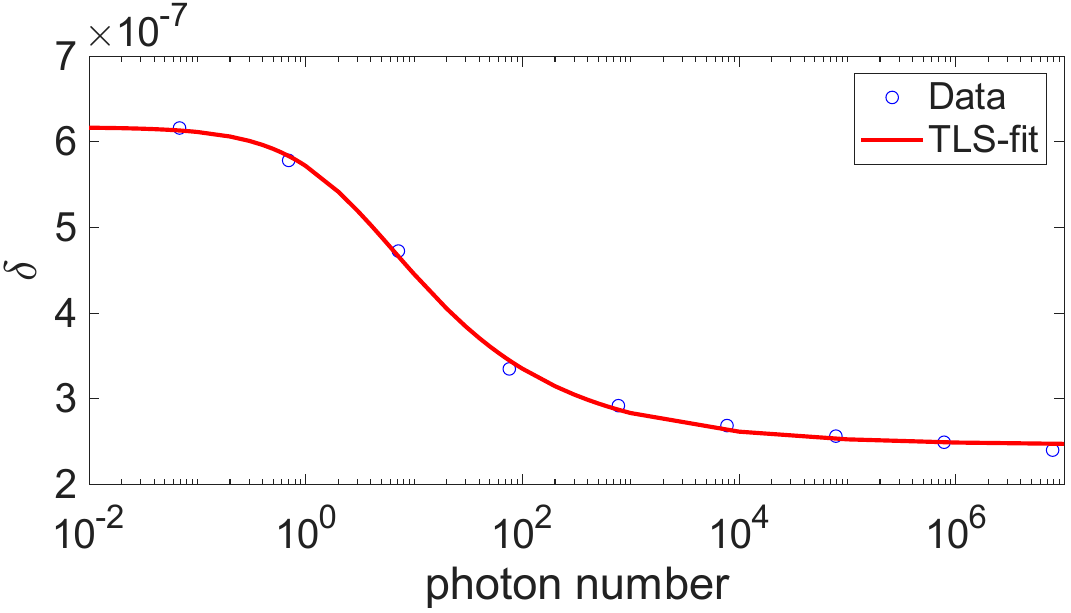}}
\caption{Exemplary plot of $\delta$ vs.\ photon number with measured data (blue) from a 4.36\,GHz resonator and a fit using Eq.~(1) (red).}
\label{fig:resonator_fit}
\end{figure}

The measurement procedure is as follows: after locating the resonance peaks with a coarse transmission sweep, the complex scattering parameter $S_{21}$ is recorded. Both are done using a vector network analyzer (VNA) \cite{simons_coplanar_2001, goppl_coplanar_2008}, which is capable of generating and analyzing microwave signals. The internal quality factor $Q_\mathrm{i}$ is then extracted by fitting the data according to Ref.~\cite{probst_efficient_2015}. Using the method described in \cite{bruno_surface_2015}, the VNA excitation power can be converted into photon number. The internal quality factor, defined as the inverse material loss $Q_\mathrm{i} = 1/\delta$, can be expressed in terms of the dielectric loss \cite{martinis_decoherence_2005}
\begin{equation}
\delta = 
\delta_\mathrm{TLS} \,
\frac{1}{\left(1 + \langle n \rangle / n_\mathrm{c}\right)^\beta}
+ \delta_\mathrm{HP},
\end{equation}
assuming a broad distribution of TLS within continuum approximation and $T \rightarrow 0$. Here, $\langle n\rangle$ is the photon number, $n_\mathrm{c}$ the critical photon number, $\beta$ an exponent, $\delta_\mathrm{TLS}$ the TLS loss and $\delta_\mathrm{HP}$ the high-power loss. Since qubits operate at n $\sim$ 1, we target the single‑photon regime, where intrinsic TLS losses dominate. Thus, we define $\delta_\mathrm{LP} = 1/Q_\mathrm{LP} = \delta(n = 1)$ and fit the data (see Fig.~\ref{fig:resonator_fit}) with Eq.~(1) to extract $\delta_\mathrm{TLS}$.

This data processing was performed for each resonator individually, while the spread of $\delta_\mathrm{LP}$ was statistically analyzed using the interquartile range (IQR) with its median $\Tilde{\delta}_\mathrm{LP}$. For the subsequent analysis, we used the low-power material loss to evaluate the suitability of our post-treatment methods for superconducting qubit applications. In general, our resonators are limited by TLS-induced losses, which are located in the different interfaces metal-air (MA), substrate-air (SA) and metal-substrate (MS). Their respective participation-ratios were extracted from FEM simulations of the CPW cross-section and are identical to \cite{lang_optimizing_2025}.

\section{Experimental}
\begin{figure}[htbp]
\centerline{\includegraphics[width=0.99\linewidth]{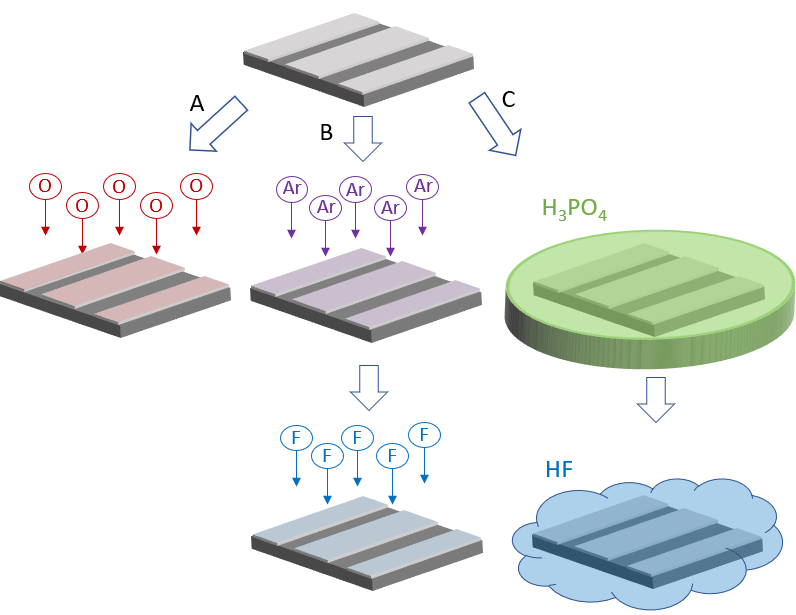}}
\caption{Schematic representation  of the different post-treatments applied to the resonator chips. Approach A utilized oxygen plasma under different conditions. In approach B, the Al surface was cleaned in-situ using Ar ion milling, followed by plasma processing with CF$_4$. In approach C, H$_3$PO$_4$ acid and HF vapor were applied both individually and in combination.}
\label{fig:schematic}
\end{figure}

The fabrication of resonator chips on 200 mm p-doped Si (100) wafers with $>$3\,k$\Omega $cm follows the same process parameters, substrate selection, metal deposition, etching and cleaning procedures as detailed in \cite{lang_CMOS_2025} for the first metal layer. Following the post-etch resist stripping by $\mathrm{H}_2\mathrm{O}$ plasma ashing, which results in the formation of thicker surface oxides compared to the native $\mathrm{SiO}_2$ and $\mathrm{Al}_2\mathrm{O}_3$, and EKC cleaning the wafer is coated with protective resist for dicing. The plasma ashing is needed to remove in-situ Cl contamination from the etching process, which degrade the Al surface. After dicing and resist removal using acetone, isopropanol and deionized water, the fabrication process for the untreated reference samples is completed. It should be noted that we obtained all samples used in this study from a single 200 mm wafer to suppress wafer-to-wafer variability and enable a more controlled and comparable evaluation of process effects.
Three different post-treatment approaches (see Fig. \ref{fig:schematic}) were applied separately on these samples, which are in principle compatible with wafer-scale implementation.

\subsection{Passivation via Oxidation}
\label{sec:ApproachA}
This approach (see Fig. \ref{fig:schematic}) aimed to compare various systems and processes of oxygen plasma. A remote oxygen plasma system with parameters akin to those reported in \cite{lang_aluminum_2023} was employed for durations of 20\,sec and 120\,sec, which enhance the oxide thickness (approximately 5\,nm \cite{lang_aluminum_2023}) and oxygen density compared to native aluminum. Similar effects are known for silicon oxides \cite{lerch_low-temperature_2009}.
Additionally, a direct oxygen plasma treatment using an Oxford Instruments PlasmaPro 80 system in RIE mode was conducted at high (290\,W) and low (50\,W) power, at 300\,mT for 3 minutes. The low-power recipe represents a standard resist stripping process well established in the EMFT cleanroom, while the high-power process was tested to improve the ashing capabilities for enhanced surface cleaning.

\subsection{Passivation via Fluorination}
\label{sec:ApproachB}
Here, we focused on passivating the aluminum surface with fluorine while preserving a pristine silicon surface (see Fig. \ref{fig:schematic}). The Oxford Instruments PlasmaPro 80 tool was employed in RIE mode due to its capability of executing the entire fabrication scheme in situ. Initially, the aluminum surface oxide was removed using Ar plasma, with process times varied for optimal oxide etching. Subsequently, CF$_4$ plasma was applied at 60\,W and 250\,mT for 3 minutes to fluorinate the exposed aluminum surface. The stability of the passivation layer was then validated through subsequent buffered oxide etching (BOE) for 30 seconds or HF vapor of an equivalent of 150\,nm SiO$_2$ in the uEtch tool from KLA, which offered high selectivity to silicon and aluminum.

\subsection{Post-Etching}
\label{sec:ApproachC}
This approach (see Fig. \ref{fig:schematic}) investigated the removal of $\mathrm{SiO}_2$ and $\mathrm{Al}_2\mathrm{O}_3$ from the surfaces. For silicon oxide removal, HF vapor was utilized again in the uEtch tool for its selectivity to silicon and aluminum \cite{Ritala_Studies_2010}. The process conditions were mentioned before (see Section~\ref{sec:ApproachB}) and the reduction of silicon oxide thickness was verified with ellipsometry measurements. For aluminum oxide etching, diluted phosphoric acid ($\mathrm{H}_3\mathrm{PO}_4$), known from conventional aluminum wet etch chemistry \cite{koehler_aetzverfahren_1998}, was employed due to its high selectivity for silicon/SiO$_2$ \cite{VanGelder_Etching_1967, Williams_Etch_1996}. In our methodology, we increased the concentration of the solution to maintain high selectivity for aluminum \cite{koch_fotolithographie_2017}. At a etch time of 1 minute, we were able to observe a saturation of the remaining aluminum oxide thickness with ellipsometry, which we attribute to the regrowth of native oxide in ambient air. Both treatments were applied to samples individual as well as in combination.

Upon completion of the post-treatment, a 24-hour coupling period begins until vacuum is reestablished during cooldown in the cryostat. During this time frame, the samples were stored in a vacuum desiccator at a pressure below 35\,$mbar$ and were retrieved only for bonding onto the PCB and mounting in the cryostat. The measurement of $\delta_{LP}$ finalizes the characterization process.

\section{Results and Discussion}
All chips used in this study were taken from the same wafer, implying identical SM across devices. The resulting metal layer thickness and linewidth variations are comparable to \cite{lang_CMOS_2025} and the samples can therefore be treated as geometrically identical. Each post-treatment described in this work was applied to two resonator chips, which were subsequently characterized in the dilution refrigerator. Some chips were lost due to handling errors, which is reflected by fewer data points in the figures. All chips were characterized using identical measurement procedures as an effort to distinguish between losses related to SA and MA interfaces. By analyzing the $\delta_\mathrm{LP}$ of the resonators, we benchmark the suitability of each process for superconducting qubit applications.

\subsection{Passivation via Oxidation}

\begin{figure}[htbp]
\centerline{\includegraphics[width=0.95\linewidth]{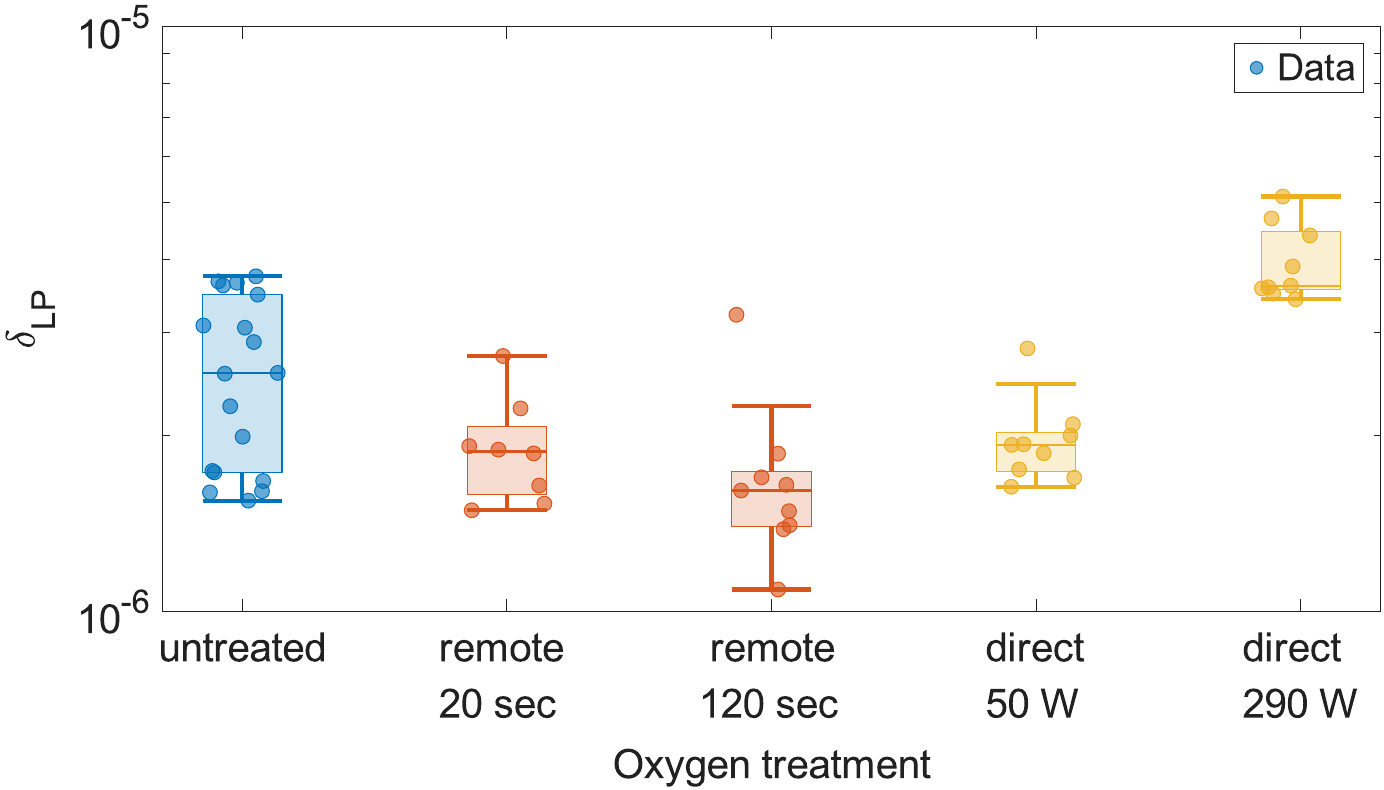}}
\caption{Low-power loss $\delta_\mathrm{LP}$ of Al resonators after different oxygen plasma treatments.}
\label{fig:oxygen}
\end{figure}

For an initial assessment of the process influence on material loss, oxygen plasma treatments were performed in remote and direct mode (see Section~\ref{sec:ApproachA}). As MA and SA surfaces are affected simultaneously, we can not distinguish individual losses.  Remote oxygen plasma slightly reduced $\delta_\mathrm{LP}$ compared to untreated samples, with longer exposure times yielding marginally lower losses (see Fig.~\ref{fig:oxygen}). Overall, $\Tilde{\delta}_\mathrm{LP}$ for the 120\,s remote plasma process showed a reduction of approximately 50\% compared to untreated samples, although the IQR remain relatively large and partially overlapping. This is in contrast to others \cite{oconnell_microwave_2008,cicak_low-loss_2010}, which reported on increased dielectric losses resulting from thicker oxides. The reduced loss may be attributed to oxygen radicals filling oxygen vacancies in the non-stoichiometric surface oxides or to additional ashing of polymer residues originating from the protective resist. The 50\,W direct plasma process yielded similar results to the 20\,sec remote plasma.

However, the high-power (290\,W) direct plasma process increased  $\Tilde{\delta}_\mathrm{LP}$ beyond the upper quartile of the untreated reference. The elevated ion energies likely damage both the Si and Al surfaces, thereby increasing the dielectric losses of the MA and SA interface, making this process unsuited for our post processing purpose.


\subsection{Passivation via Fluorination}

CF$_4$ plasma treatments on Al surfaces were investigated as a potential passivation technique. Since CF$_4$ plasmas can either etch or deposit fluorocarbon films depending on process parameters \cite{lim_comparison_2021, cho_expression_1999}, a short parameter study with varying plasma powers was carried out. Based on this, a CF$_4$ process was selected that incorporated fluorine into the Al surface without depositing C–F residues on the Si surface (see Section~\ref{sec:ApproachB}).

\begin{figure}[htbp]
\centerline{\includegraphics[width=0.9\linewidth]{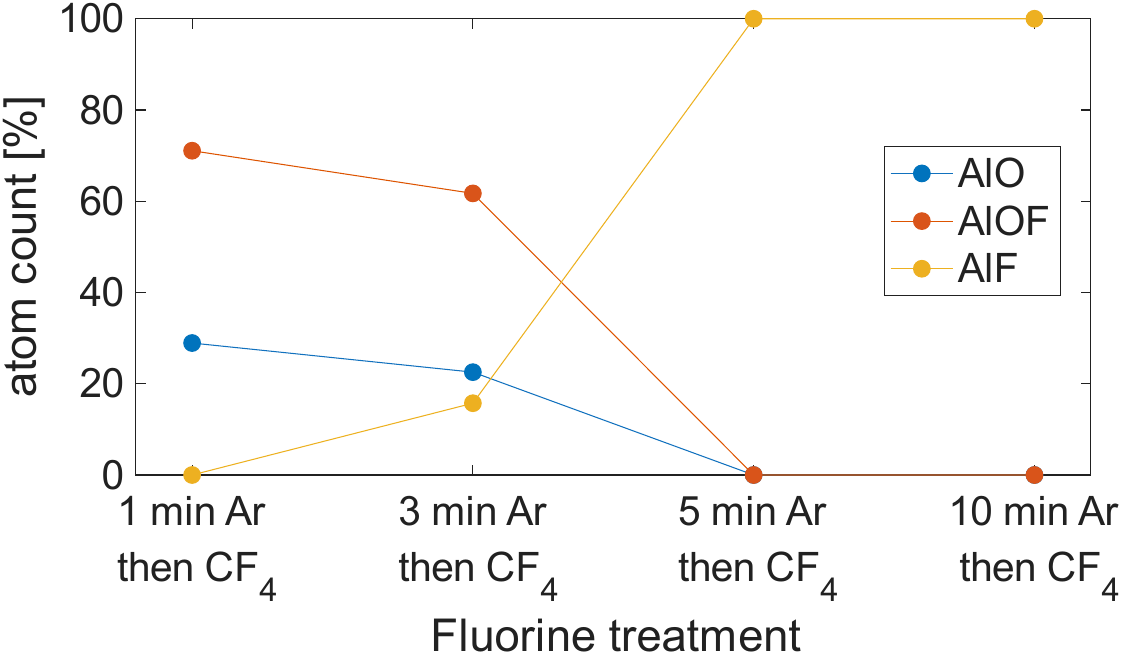}}
\caption{Bond concentration versus Ar ion milling duration prior to CF$_4$ plasma treatment. The surface was cleaned for 1\,min using Ar ion milling before the XPS measurements were carried out. Lines are for guiding the eyes.}
\label{fig:flourine_XPS}
\end{figure}

To efficiently remove the native Al$_2$O$_3$ layer prior to fluorination, Ar ion milling with different durations was investigated. Figure~\ref{fig:flourine_XPS} shows the XPS-derived relative bonding concentration of Al--O, Al--F and Al--O--F for different Al surface preparation conditions. The calculation of the concentrations is explained in Section \ref{sec:methodic}. Increasing ion-milling duration results in a higher concentration of F-containing bonds by reducing oxygen bond concentration, reaching saturation after approximately 5\,min, with no Al--O or Al--O--F bonds left on the surface. The incorporation of fluorine into the surface is energetically favored ($\mathrm{G}_{\mathrm{AlF}} = 675$\,kJ/mol vs.\ $\mathrm{G}_{\mathrm{AlO}} = 502$\,kJ/mol \cite{rumble_crc_2022}), supporting the approach of reducing oxygen through fluorination. Initially, Ar ion milling is removing the native aluminum surface oxide only partially, allowing a remaining oxygen concentration on the surface. Increased Ar milling duration is etching into the Al and provides a oxygen-free surface solely covered with Al--F bonds after CF treatment. As we can identify the absence of O-containing bonds on this surface in the XPS measurements (after several days of delay), a long-lasting passivation with fluorine was established.

\begin{figure}[htbp]
\centerline{\includegraphics[width=0.95\linewidth]{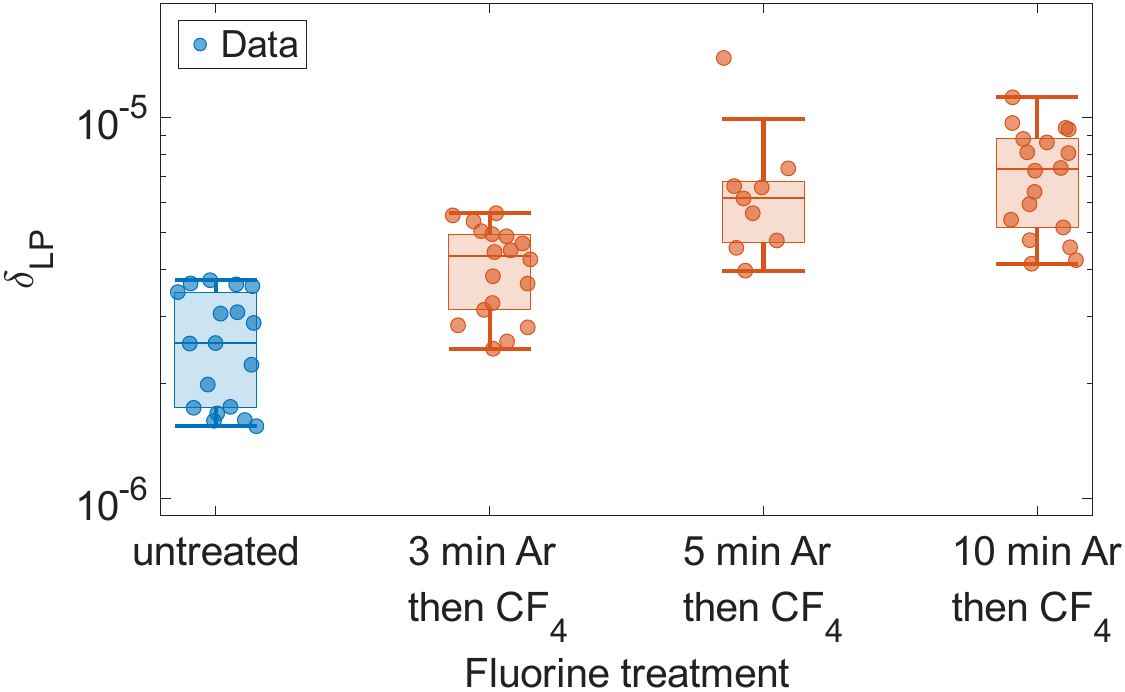}}
\caption{Low-power loss $\delta_\mathrm{LP}$ of fluorinated samples for different Ar ion-milling durations.}
\label{fig:flourine}
\end{figure}

These fluorination processes were subsequently applied to resonator chips and characterized at millikelvin temperatures. As shown in Fig.~\ref{fig:flourine}, the low-power loss of our milled and flourinated samples increased compared to untreated samples. A substantial increase in dielectric loss due to Ar milling is not expected, as previous studies \cite{van_damme_argon-milling-induced_2023, quintana_characterization_2014} already demonstrated. However, a minor increase in loss caused by the Ar ion milling is evident between the 5\,min ($\Tilde{\delta}_\mathrm{LP} = 6.1\times10^{-6}$)  and 10\,min ($\Tilde{\delta}_\mathrm{LP} = 7.3\times10^{-6}$)  samples, as the fluorine concentration is identical. This is supported by AFM measurements of the silicon surface after the 5\,min (R$_\mathrm{RMS}$ = 1.1\,nm) and 10\,min (R$_\mathrm{RMS}$ = 1.2\,nm) Ar milling, which increase in roughness compared to the untreated surface (R$_\mathrm{RMS}$ = 0.7\,nm). Scaling this increase in loss by Ar-ion milling to the untreated sample suggests that the incorporation of fluorine in the surface layer did not substantially increase the LP (low-power) loss of our resonators.

\begin{figure}[h]
	\vspace{6mm}
	\centering
	\begin{subfigure}[b]{\linewidth}
		\centering
		\begin{overpic}[width=0.95\textwidth]{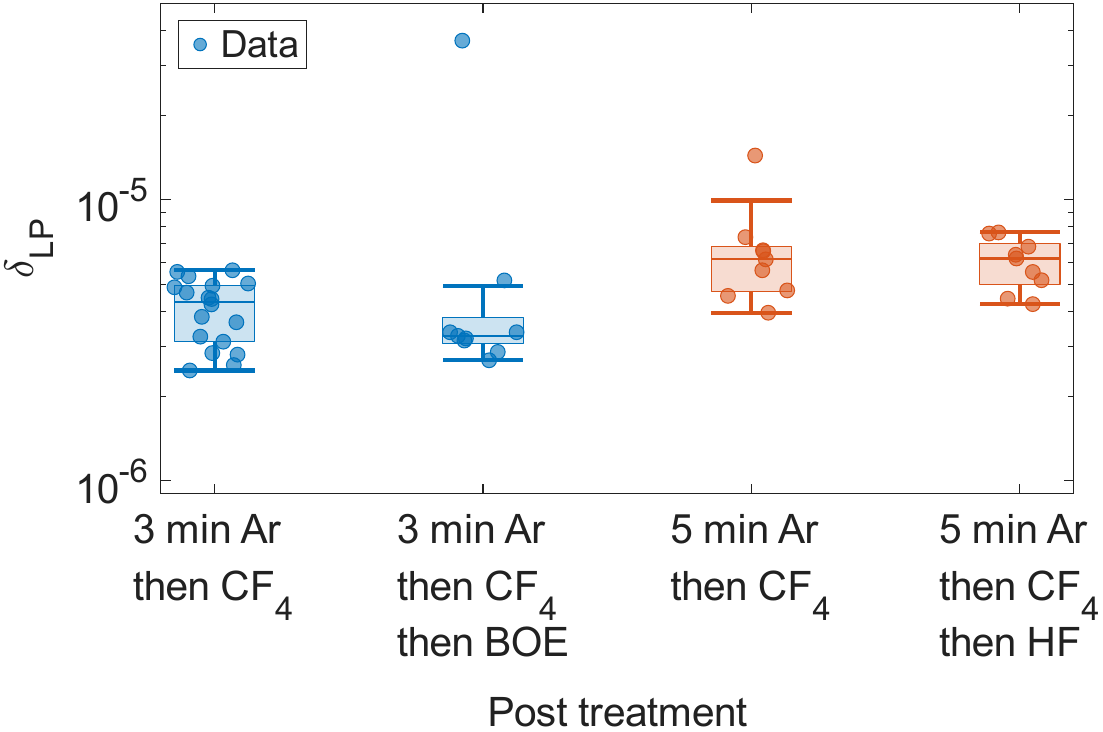}
		\put(-1,68){\textbf{(a)}}
		\end{overpic}
	\end{subfigure}
	\vspace{2mm}
	\begin{subfigure}[b]{\linewidth}
		\centering
		\begin{overpic}[width=0.95\textwidth]{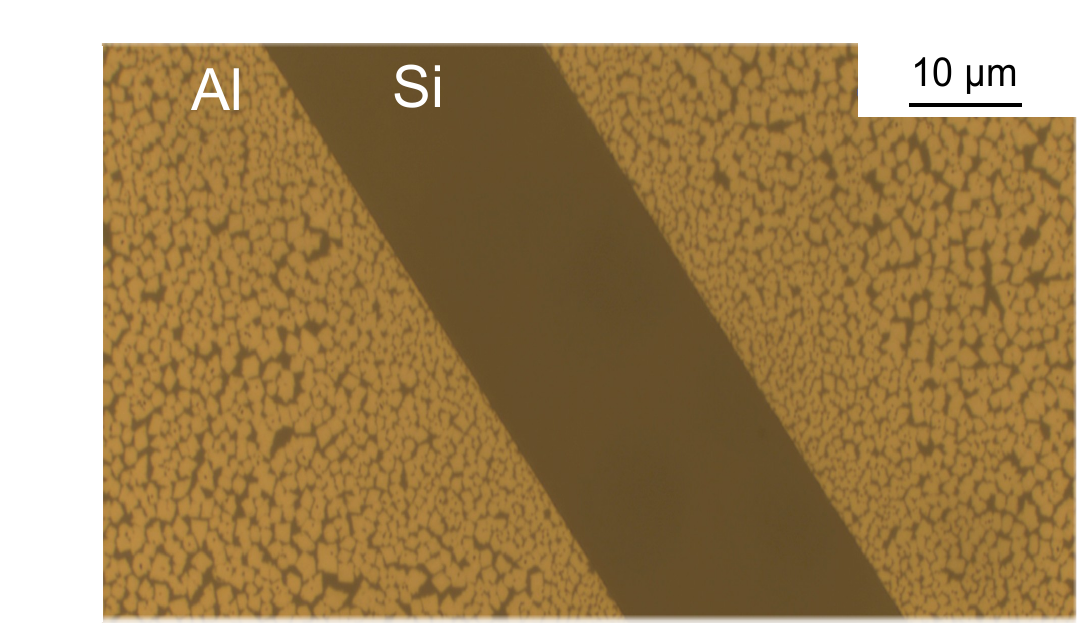}
		\put(-1,55){\textbf{(b)}}
		\end{overpic}
	\end{subfigure}
    \caption{(a) Effect of BOE and HF post-etching on fluorinated Al resonators. (b) If no passivation is apparent, the Al surface is visibly attacked by 30\,sec BOE.}
    \label{fig:flourine_post-etch}
\end{figure}

To evaluate whether Ar milling induced losses to the Si surface by trench damage or surface amorphization \cite{marcaud_low-loss_2025, lang_optimizing_2025} can be removed, additional post-etching using BOE or HF vapor was performed. For the 3\,min milled sample (see Fig.~\ref{fig:flourine_post-etch},a), BOE did not substantially reduce $\Tilde{\delta}_\mathrm{LP} = 3.3\times10^{-6}$ compared to the non-etched sample $\Tilde{\delta}_\mathrm{LP} = 4.3\times10^{-6}$, as both IQR nearly overlap. However, the median reduced by about 23\% and is therefore moving closer to the untreated sample with $\Tilde{\delta}_\mathrm{LP} = 2.6\times10^{-6}$. Removed losses from the Si surface have to be be considered as a reason. Interestingly, BOE did not visibly damage the Al surface, which would normally be expected for untreated Al (see Fig.~\ref{fig:flourine_post-etch}b), where the BOE etches along the grain boundaries until the Si substrate is reached, resulting in the collapse of superconductivity in the Al layer.  This suggests that the fluorinated surface is sufficiently passivated to withstand BOE, with a Al--F bonding concentration as low as 15.7\% appearing sufficient. HF vapor treatment, however, did not lead to a reduction in loss as $\Tilde{\delta}_\mathrm{LP}$ with and without HF are comparable. Compared to the BOE, the HF vapor is less likely to remove Ar milling induced losses on the Si surface.

Above all, these findings suggest that fluorine incorporation stabilizes the Al surface and passivates it against aggressive Al/Al$_2$O$_3$ etchants like BOE. However, for our proposed treatment scheme the low-power loss in the superconducting regime increases by a factor of 2.4 - 2.9 for purely Al--F covered surfaces.

\subsection{Post-Etching}
\begin{figure}[htbp]
\centerline{\includegraphics[width=0.95\linewidth]{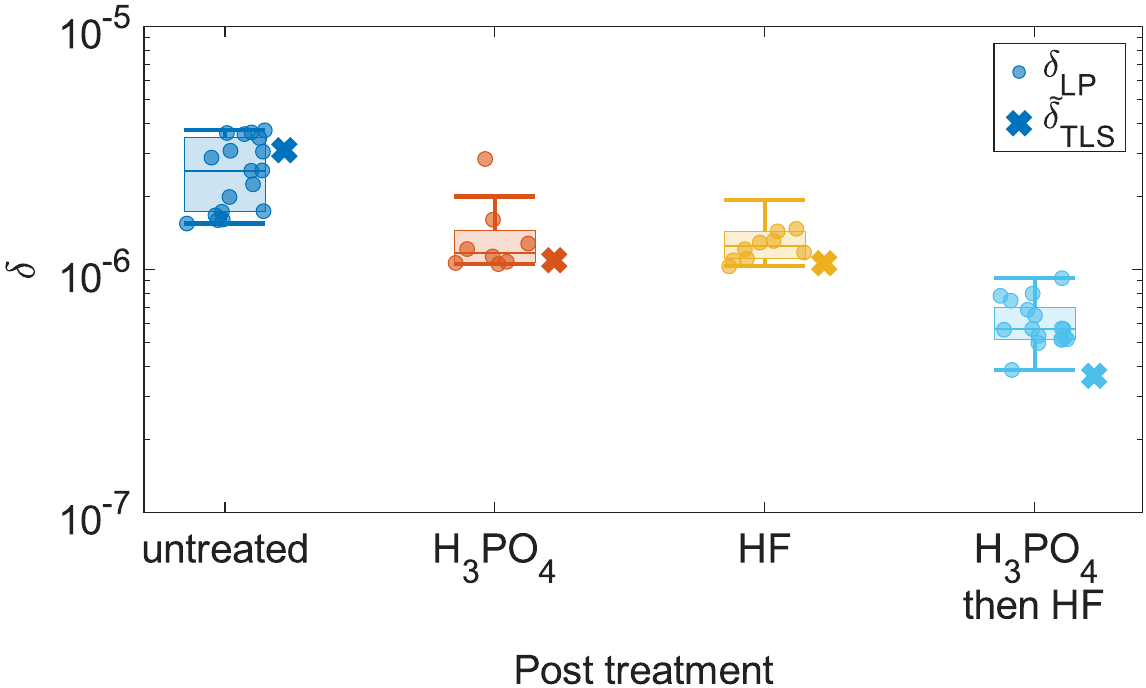}}
\caption{Low-power loss $\delta_\mathrm{LP}$ and $\Tilde{\delta}_\mathrm{TLS}$  after HF vapor etch, phosphoric acid, and combined post-etching treatments.}
\label{fig:pure_post}
\end{figure}

As a final post-processing approach, HF vapor and phosphoric acid were used to selectively remove SiO$_2$ and Al$_2$O$_3$ from the resonator surfaces (see Section~\ref{sec:ApproachC}) of untreated samples. The chips were subsequently cooled down for cryogenic characterization. Compared to untreated samples, the phosphoric acid reduced $\Tilde{\delta}_\mathrm{LP}$ by approximately a factor of 2 (see Fig.~\ref{fig:pure_post}), separating both IQR. We attribute this improvement primarily to reduced MA loss by aluminum oxide thickness reduction, owing to the chemical nature of phosphoric acid. Analogously, HF vapor etching achieved similar improvements by reducing silicon oxide thicknesses at the SA surface. Applying both etches sequentially further reduced the loss by a combined factor of three, resulting in $\Tilde{\delta}_\mathrm{LP} = 5.7 \times 10^{-7}$ with TLS-related losses approaching $\Tilde{\delta}_\mathrm{TLS} = 3.6 \times 10^{-7}$. This corresponds to a quality factor of $Q_\mathrm{LP} = 1.7$\,M and $Q_\mathrm{TLS} = 2.8$\,M, which represents the highest values observed throughout this study. These findings support the well-established idea that surface oxides host TLS defects, which increase low-power losses~\cite{martinis_decoherence_2005, muller_towards_2019, gao_experimental_2008}. Notably, we were able to effectively reduce oxide-related losses despite a 24\,h delay between fabrication and cooldown. This substantial reduction may result from increased oxide thickness and losses in the MA and SA surfaces caused by the water vapor plasma resist removal, which may not be as detrimental for wet etch aluminum resonators without H$_2$O plasma strip.

\begin{figure}[htbp]
\centerline{\includegraphics[width=0.95\linewidth]{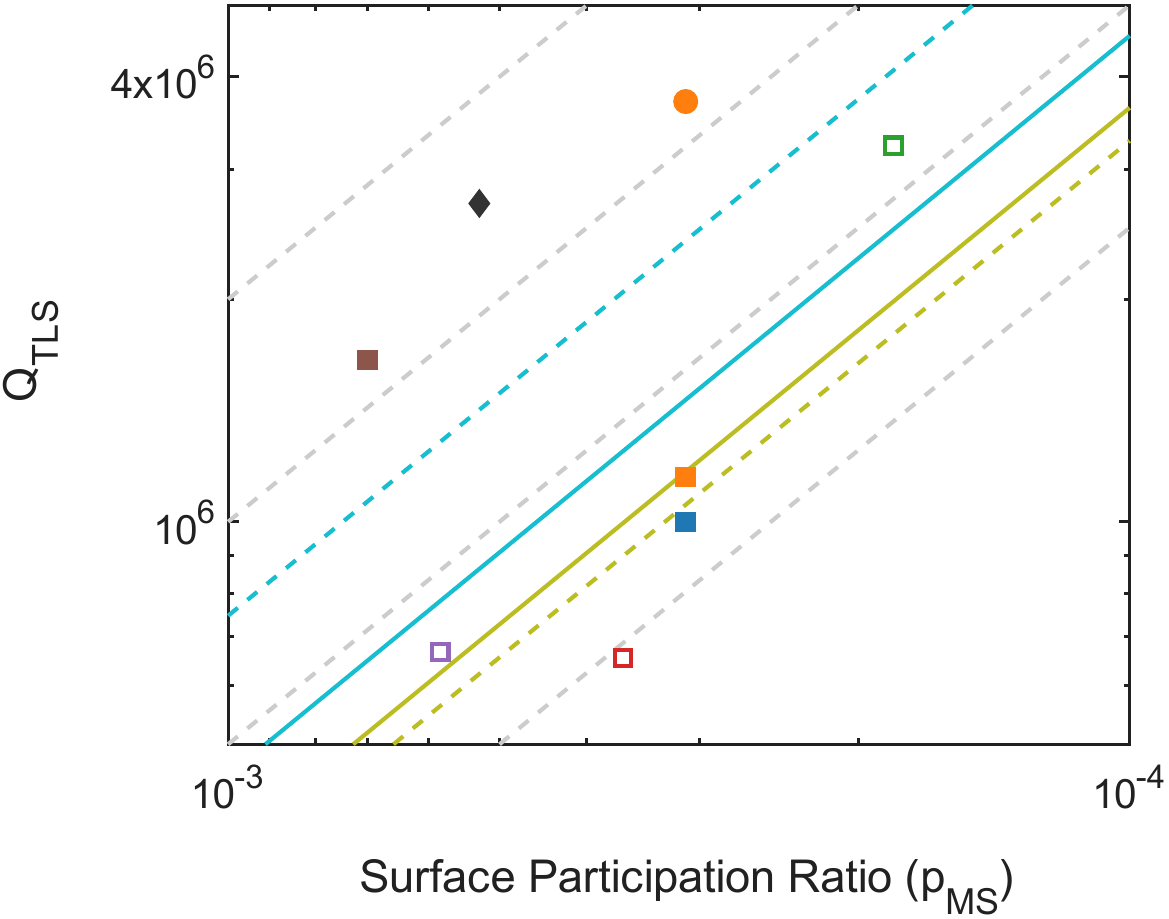}}
\caption{TLS related quality factor against the surface participation ratios of resonators calculated analytically with \cite{murray_analytical_2018} for various published wet or dry etched aluminum-on-Si resonators with an over-etch of less then 50\,nm (squares or solid lines) and additional post-treatments (circles or dashed lines). For the lines, dielectric losses calculated from resonators were used. The grey dashed lines indicate constant TLS-related losses scaling linear with $p_\mathrm{MS}$. Open marker resemble single resonator, filled mean values over a chip with several resonators. Black diamond: our phosphoric and HF vapor treated resonators. Extracted data from publications: light-blue \cite{hedrick_quantifying_2026}, light-yellow \cite{melville_comparison_2020}, green \cite{dunsworth_characterization_2017}, blue \cite{biznarova_mitigation_2024}, brown \cite{zikiy_high-q_2023}, red \cite{earnest_substrate_2018, mcrae_materials_2020}, purple \cite{sage_study_2011, mcrae_materials_2020}, orange \cite{chayanun_characterization_2024}.  }
\label{fig:overview_QTLS}
\end{figure}

High TLS-related quality factors alone do not necessarily indicate a superior fabrication process, as they are device-specific and therefore challenging to relate across different benchmark processes. To generalize our results in the context of aluminum-on-Si based resonators, Figure~\ref{fig:overview_QTLS} compares the TLS-related quality factors for different resonator geometries by using their surface participation ratios. Here, we used an analytical approach \cite{murray_analytical_2018} to calculate $p_\mathrm{MS}$. Along the light-grey dashed lines, the limiting TLS-losses are equal and scale therefore with the surface participation. Other approaches used (diluted) liquid HF \cite{hedrick_quantifying_2026, melville_comparison_2020, chayanun_characterization_2024} (Fig.~\ref{fig:overview_QTLS}, light-blue and light-yelow lines, orange markers) with shorter duration and less selectivity \cite{Williams_Etch_1996} to remove oxides on Si and Al for reduced losses. In comparison, our treatment based on selective etching using HF vapor and phosphoric acid is more reproducible due to longer etch duration and, to the best of our knowledge, achieves the lowest published $\delta_\mathrm{TLS}$ for the given surface participation ratio. Moreover, the data from the present study (Fig.~\ref{fig:overview_QTLS}, black diamond) incorporate chip-to-chip variation with
median values over two chips with 9 resonator each. Interestingly, wet-etched resonators exhibit only slightly higher losses \cite{zikiy_high-q_2023} (Fig.~\ref{fig:overview_QTLS}, brown square) then our post-treatment, which supports the idea of vapor plasma as reason for the initially high MA and SA losses of the reference resonators. 

\section{Conclusion}
In this study, several strategies were investigated to post-treat the surfaces of aluminium resonators prior to a cooldown occurring 24\,h after processing. Oxygen and fluorine plasmas were used as passivation approaches. Plasma oxidation marginally reduced $\delta_\mathrm{LP}$ when low-power direct plasma or remote plasma was applied, which is likely attributed to additional ashing that removes residual resist or polymer contamination. In contrast, high-power oxygen plasma increased the losses, which can be caused by the surface impact of high energetic ions.

Fluorine plasma treatment in combination with Ar ion milling enables to fully cover Al surfaces with Al--F bond with increasing milling duration, which correlated with an increase in $\Tilde{\delta}_\mathrm{LP}$ of about 2.4 - 2.9. The fluorine-treated surface exhibited a passivating effect against BOE etching with Al--F concentrations starting 15.7\%, enabling controlled preparation of the aluminium metallization for subsequent BOE post-treatments. This is particularly relevant for device architectures in which Nb or Ta are used in combination with Al. The fluorinated Al surfaces, including Josephson junction electrodes, can be sufficiently stabilized to allow post-BOE dipping of the entire chip, which is crucial to achieve high quality factors with Nb resonators. 

An estimation should be given as follow: A typical TLS-related qubit loss is given by $\delta_\mathrm{qubit}p^\mathrm{qubit}_\mathrm{MS}  = \delta_\mathrm{pads} p^\mathrm{pads}_\mathrm{MS} + \delta_\mathrm{junction} p^\mathrm{junction}_\mathrm{MS} $ with a pad-related participation ratio $p^\mathrm{pads}_\mathrm{MS} \sim 5.2\times 10^{-5}$ \cite{gambetta_investigating_2017, bland_2d_2025, murray_analytical_2018} and a junction participation of about $p^\mathrm{junction}_\mathrm{MS} \sim 1.1\times 10^{-5}$ \cite{murray_analytical_2018}. We focus on the surface participation ratio ($p_\mathrm{MS}$) similar to previous studies \cite{crowley_disentangling_2023,wang_surface_2015,hedrick_quantifying_2026}. Assuming a fabrication scheme where Nb resonators can be protected from the passivation treatment with resist or sacrificial oxides and a dominant junction loss similar to the aluminum loss presented in this study the increased loss of the junction by the mildest fluorine treatment $\delta^{F}_{junction} / \delta_{junction} = 1.7 $ is compensated by the decrease in loss of untreated Nb resonators ($\delta_{pads} = 0.81\times 10^{-6} $ on average \cite{verjauw_investigation_2021, altoe_localization_2022}) through the subsequent BOE treatment to about $\delta^{BOE}_{pads} / \delta_{pads} = 0.25$ on average \cite{verjauw_investigation_2021, altoe_localization_2022}. This offers reduction in qubit losses of $\delta^{F+BOE}_{qubits} / \delta_{qubits} = 0.9$. However, others \cite{lisenfeld_mapping_2025} demonstrated that about 59\% of strong coupling TLS are found near the junction, which are not taken in account by our assumptions. 

Finally, selective post-etching with vapor HF and phosphoric acid was employed to remove SiO$_2$ and Al$_2$O$_3$ from the resonator surfaces. Using both etches sequentially reduced the dielectric losses substantially compared to the untreated samples, by a factor of three, down to $\Tilde{\delta}_\mathrm{TLS} = 3.6 \times 10^{-7}$ or $Q_\mathrm{TLS} = 2.8$\,M and is therefore very suitable for the post processing of superconducting Al-based qubits. Using the analytical approach \cite{murray_analytical_2018} also allows us to directly estimate the coherence time $T_1$ for given qubit geometries from our results. As mentioned before, typical qubit designs offer $p_\mathrm{MS} \approx 5.2\cdot10^{-5}$, which results in $T_1 \approx 1.2$\,ms for a target frequency of 4\,GHz assuming no additional limiting loss.
In this regard, our resonator-based robust fabrication scheme demonstrate both the relevance and the feasibility of targeted post-treatments on single chips for quantum devices, even within an industrial fabrication environment.

\section*{Acknowledgements}
The authors acknowledge helpful discussions with G. Huber, I. Tsitsilin, F. Haslbeck, C. Schneider, L. Koch and N. Bruckmoser from the Quantum Computing group at the Walther Meissner Institute. The authors thank A. Cleland from the University of Chicago and H. Yan from Applied Materials for their insights and productive discussions. Additionally, we thank A. Reinholdt and K. Langenbrink from Fraunhofer ISC for the XPS measurements and Z. Luo for his support with simulation. We also appreciate M. Hahn and M. König for their help and valuable discussions in process development and the whole Fraunhofer EMFT clean room staff for the professional fabrication, especially Luca Rommeis.
\\ \\
This work was funded by the Munich Quantum Valley (MQV) – Consortium Scalable Hardware and Systems Engineering (SHARE), funded by the Bavarian State Government with funds from the Hightech Agenda Bavaria, the Munich Quantum Valley Quantum Computer Demonstrator - Superconducting Qubits (MUNIQC-SC) 13N16188, funded by the Federal Ministry of Education and Research, Germany, and the Open Superconducting Quantum Computers (OpenSuperQPlus) Project - European Quantum Technology Flagship.

\bibliography{paper.bib}

\end{document}